\documentclass[final,twocolumn,showpacs,amsmath]{revtex4}
\usepackage{graphicx}
\numberwithin{equation}{section}

\begin{document}

\title{On the Transport Diamonds and Zero Current Anomaly in InGaAs/InP and GaAs/AlGaAs}

\author{S.~Fujita}
\email{fujita@buffalo.edu}
\affiliation{Department of Physics, SUNY at Buffalo, Buffalo, New York 14260, USA}

\author{H.\,C.~Ho}
\email{hcho@sincerelearning.hk}
\affiliation{Sincere Learning Centre, Kowloon, Hong Kong SAR, China}

\date{\today}

\begin{abstract}
In the quantum Hall effect (QHE) the differential resistivity $r_{xx}
\equiv r$ vanishes within a range where the Hall resistivity forms a
plateau. A microscopic theory is developed, starting with a crystal
lattice, setting up a BCS-like Hamiltonian in terms of composite bosons,
and using statistical mechanical method. The main advantage of our
bosonic theory is its capability of explaning the plateau formation in
the Hall resistivity, which is assumed in the composite fermion
theories. In the QHE under radiation, the resistivity vanishes within a
range with no plateau formation. This is shown in terms of two-channels
model, one channel excited by radiation where the supercurrents run and
the other (base) channel in which the normal currents run. The transport
diamonds (TD) and the zero direct current anomaly (ZCA) occur when the
resistivity $r$ is measured as a function of magnetic field and direct
current (DC). The spiral motion of an electron under a magnetic field
can be decomposed into two, the cyclotron motion with the cyclotron mass
$m^*$ and the guiding center motion with the magnetotransport $M^*$. The
quantization of the motion generates magnetic oscillations in the
density of states. The magnetoconductivity is calculated, using kinetic
theory and quantum statistical mechanics. The TR and ZCA are shown to be
a breakdown of QHE. The integer QHE minima are shown to become the
Shubnikov-de Haas (SdH) maxima progressively as the DC increases. The
ZCA at low temperatures ($T=0.253$--$1.2$\,K) is temperature-dependent,
which is caused by the electron-optical-phonon scattering.
\end{abstract}

\pacs{73.43.-f, 73.43.Qt, 74.25.Ha}

\maketitle

\section{INTRODUCTION}\label{se:Introduction}

Recently, Studenikin \emph{et~al}.\ \cite{Studenikin} discovered
transport diamonds (TR) and zero-current anomaly (ZCA) in
In$_x$Ga$_{1-x}$As/InP and GaAs/AlGaAs. The TR and ZCA occur when the
differential resistance $r_{xx} = dV_{xx}/dI$ of a Hall bar sample after
the red-light illumination is plotted in the plane of the magnetic field
and the direct current (DC). See Fig.~\ref{fg:Figure4}, which is reproduced from
Ref.~\cite{Studenikin}, Fig.~2. Diamond-shaped regions developing from
SdH minima are called \emph{transport diamonds} and a sharp dip in
$r_{xx}$ appearing at a narrow horizontal line at zero DC is called a
\emph{zero-current anomaly}. The details of the experiments and
theoretical backgrounds can be found in Ref.~\cite{Studenikin}. The
original authors~\cite{Studenikin} suggested an interpretation: the
breakdown of the quantum Hall effect (QHE). We shall show in the present
work that this is indeed the case based on the composite (c-)boson
model, the model originally introduced by Zhang, Hansson and
Kivelson~\cite{Zhang} and later developed by Fujita's
group~\cite{Fujita_1}. In the prevalent theories \cite{Ezawa}, the QHE
is discussed in terms of the c-fermions. \cite{Jain} The formation of
the Hall resistivity plateau where the resistivity vanishes, is assumed,
however. In our c-boson theory, the plateau formation is explained from
the first principles. In Sec.~\ref{se:Integer_QHE}, we review our microscopic theory of
the QHE. The QHE for a system subjected to a radiation is developed and
discussed in Sec.~\ref{se:QHE_Radiation}. The density of Landau states and statisical
weight in two dimensions (2D) are calculated in Sec.~\ref{se:DOS_SW}. Shubnikov-de
Haas (SdH) oscillations \cite{Shubnikov} in the magnetoconductivity and de
Haas-van Alphen (dHvA) oscillations \cite{Haas} in the magnetic
susceptibility, are jointly called magnetic oscillations. They originate
in the oscillations in statistical weight. Magnetic oscillations are
often discussed, using a so-called Dingle temperature \cite{Dingle}. But
this is a phenomenological treatment, which must be avoided. Following
our previous work \cite{Fujita_6}, we present a microscopic theory of
the SdH oscillations for a 2D system in Sec.~\ref{se:SdH_Oscillations}. The TR and ZCA are
discussed in Sec.~\ref{se:TD_ZCA}.

\section{INTEGER QUANTUM HALL EFFECT}\label{se:Integer_QHE}

If a magnetic field $\mathbf{B}$ is applied slowly, then the classical
electron can continuously change from the straight line motion at zero
field to the curved motion at a finite $B$ (magnitude). Quantum
mechanically, the change from the momentum state to the Landau state
requires a perturbation. We choose for this perturbation the phonon
exchange attraction between the electron and the \emph{fluxon}
(elementary magnetic flux). Consider an electron with a few fluxons. If
the magnetic field is applied slowly, the energy of the electron does
not change but the cyclotron motion always acts so as to reduce the
magnetic fields surrounding the electron. Hence the total energy of the
composite of an electron and fluxon is less than the electron energy
plus the unperturbed field energy. In other words, the composite
(c-)particle is stable against the break-up, and it is in a bound
(negative energy) state. The c-particle is simply a dressed electron
carrying $Q$ fluxons. $Q=1,2,\dots$. Originally, the c-particle was
introduced as a composite of one electron attached with a number of
Chern-Simons gauge objects. \cite{Ezawa} These objects are neither
bosons nor fermions, and hence the statistics of the composite is not
clear. The basic particle property (countability) of the fluxons is
known as the flux quantization, see Eq.~\eqref{eq:mfi_quantized}. We assume that the
fluxon is an elementary fermion with zero mass and zero charge, which is
supported by the fact that the fluxon, the quantum of the magnetic field
$\mathbf{B}$, cannot disappear at a sink unlike the bosonic photon, the
quantum of the electric field $\mathbf{E}$ \cite{Fujita_1}. Fujita and
Morabito \cite{Fujita_2} showed that the center-of-mass (CM) of a composite moves
following the Ehrenfest-Oppenheimer-Bethe's (EOB) rule \cite{Ehrenfest}: the
composite is fermionic (bosonic) if it contains an odd (even) number of
elementary fermions. Hence the quantum statistics of the c-particle is
established.

At the Landau level (LL) occupation number, also
called the \emph{filling factor}, $\nu = 1 / Q$, $Q$ odd, the c-bosons with $Q$
fluxons are generated and can condense below certain critical temperature
$T_c$. The Hall resistivity plateau is caused by kind of the Meissner
effect as explained later.

We develop a theory for GaAs/AlGaAs heterojunction, the theory which can also be applied to
InGaAs/InP quantum well. GaAs forms a zincblende lattice. We assume that
the interface is in the plane $(001)$. The Ga$^{3+}$ ions form a square
lattice with the sides directed in $[110]$ and $[1\bar{1}0]$. The
``electron'' (wave packet) with a negative charge $-e$ will then move isotropically with an
effective mass $m_1$. The As$^{3-}$ ions also form a square lattice at a
different height in $[001]$. The ``holes'', each having a positive
charge (+e), will move similarly with an effective mass $m_2$. A longitudinal
phonon moving in $[110]$ or in $[1\bar{1}0]$ can generate a charge
(current) density variation, establishing an interaction between the
phonon and the electron (fluxon). If one phonon exchange is considered
between the electron and the fluxon, a second-order perturbation
calculation establishes an effective electron-fluxon interaction
\begin{equation}
\left|V_q V_q'\right| \frac{\hbar \omega_q}{\left(\varepsilon_{|\mathbf{p}\boldsymbol{+}\mathbf{q}|}-\varepsilon_p\right)^2 -
(\hbar\omega_q)^2},
\label{eq:aee_interaction}
\end{equation}
where $\mathbf{q}$ ($\hbar \omega_q$) is the phonon momentum (energy),
$V_q(V_q')$ the interaction strength between the electron (fluxon) and
the phonon. If the energies
$\left(\varepsilon_{|\mathbf{p}\boldsymbol{+}\mathbf{q}|}, \varepsilon_{p}\right)$ of the final and initial electron
states are equal, the effective interaction is negative (attractive) as seen from Eq.~\eqref{eq:aee_interaction}.

Following Bardeen-Cooper-Schrieffer (BCS)~\cite{Bardeen}, we start with a
Hamiltonian $H$ with the phonon variables eliminated:
\begin{align}
H = &\sum_k \sum_s \varepsilon_k^{(1)} n_{\text{k}s}^{(1)} + \sum_k
\sum_s \varepsilon_k^{(2)} n_{\text{k}s}^{(2)} + \sum_k \sum_s \varepsilon_k^{(3)} n_{\text{k}s}^{(3)} \nonumber\\
&\mbox{} - v_0 \sideset{}{'}\sum_\mathbf{q} \sideset{}{'}\sum_\mathbf{k} \sideset{}{'}\sum_{\mathbf{k}\boldsymbol{'}} \sum_s
\Big[ B_{\mathbf{k}\boldsymbol{'}\mathbf{q}s}^{(1)\dagger}B_{\mathbf{kq}s}^{(1)} +
  B_{\mathbf{k}\boldsymbol{'}\mathbf{q}s}^{(1)\dagger}B_{\mathbf{kq}s}^{(2)\dagger} \nonumber\\
&\mbox{} + B_{\mathbf{k}\boldsymbol{'}\mathbf{q}s}^{(2)}B_{\mathbf{kq}s}^{(1)}
+
B_{\mathbf{k}\boldsymbol{'}\mathbf{q}s}^{(2)}B_{\mathbf{kq}s}^{(2)\dagger} \Big],
\label{eq:tpv_eliminated}
\end{align}
where $n_{\mathbf{k}s}^{(j)}$ is the number operator for the ``electron''(1) [``hole''(2),
  fluxon (3)] at momentum $\mathbf{k}$ and spin $s$ with the energy $\varepsilon_{ks}^{(j)}$. We
represent the ``electron'' (``hole'') number $n_{\mathbf{k}s}^{(j)}$ by $c_{\mathbf{k}s}^{(j)\dagger}c_{\mathbf{k}s}^{(j)}$, where $c(c^\dagger)$ are
annihilation (creation) operators satisfying the Fermi anticommutation
rules:
\begin{align}
\left\{c_{\mathbf{k}s}^{(i)}, c_{\mathbf{k}\boldsymbol{'}s'}^{(j)\dagger}\right\}
&\equiv c_{\mathbf{k}s}^{(i)} c_{\mathbf{k}\boldsymbol{'}s'}^{(j)\dagger}
+ c_{\mathbf{k}\boldsymbol{'}s'}^{(j)\dagger} c_{\mathbf{k}s}^{(i)} =
\delta_\mathbf{k, k'} \delta_\mathbf{s, s'} \delta_{i, j}, \nonumber\\
\left\{c_{\mathbf{k}s}^{(i)}, c_{\mathbf{k}\boldsymbol{'}s'}^{(j)}\right\} &= 0.
\label{eq:Fermi_anticommutation_rules}
\end{align}
We represent the fluxon number $n_{\mathbf{k}s}^{(3)}$ by
$a_{\mathbf{k}s}^\dagger a_{\mathbf{k}s}$, with $a(a^\dagger)$, satisfying
the anticommutation rules, \eqref{eq:Fermi_anticommutation_rules}.
\begin{align}
B_{\mathbf{kq}s}^{(1)\dagger} &\equiv c_{\mathbf{k}\boldsymbol{+}\mathbf{q}/2\,s}^{(1)\dagger} a_{\boldsymbol{-}\mathbf{k}\boldsymbol{+}\mathbf{q}/2\,-s}^\dagger, \nonumber\\
B_{\mathbf{kq}s}^{(2)} &\equiv c_{\mathbf{k}\boldsymbol{+}\mathbf{q}/2\,s}^{(2)} a_{\boldsymbol{-}\mathbf{k}\boldsymbol{+}\mathbf{q}/2\,-s}.
\label{eq:anticommutation_rules}
\end{align}
The prime on the summation in Eq.~\eqref{eq:tpv_eliminated} means the
restriction: $0 < \varepsilon_{ks}^{(j)} < \hbar\omega_\text{D}$, $\omega_\text{D}$ = Debye frequency. If the fluxons are replaced by
the conduction electrons (``electrons'', ``holes'') our Hamiltonian $H$
is reduced to the original BCS Hamiltonian, Eq.~(24) of Ref.~\cite{Bardeen}. The
``electron'' and ``hole'' are generated, depending on the energy contour
curvature sign \cite{Fujita_3}. For example, only ``electrons'' (``holes''), are
generated for a circular Fermi surface with the negative (positive)
curvature whose inside (outside) is filled with electrons. Since the
phonon has no charge, the phonon exchange cannot change the net charge.
The pairing interaction terms in Eq.~\eqref{eq:tpv_eliminated} conserve the charge. The term
$-v_0
B_{\mathbf{k}\boldsymbol{'}\mathbf{q}s}^{(1)\dagger}B_{\mathbf{kq}s}^{(1)}$,
where $v_0\equiv \left|V_qV_q'\right|(\hbar\omega_0A)^{-1}$, $A$ = sample area, is the pairing strength, generates
the transition in the ``electron'' states. Similarly, the exchange of a
phonon generates a transition in the ``hole'' states, represented by
$-v_0
B_{\mathbf{k}\boldsymbol{'}\mathbf{q}s}^{(2)}B_{\mathbf{kq}s}^{(2)\dagger}$.
The phonon exchange can also pair-create and pair-annihilate
``electron'' (``hole'')-fluxon composites, represented by $-v_0 B_{\mathbf{k}\boldsymbol{'}\mathbf{q}s}^{(1)\dagger}B_{\mathbf{kq}s}^{(2)\dagger}$, $-v_0 B_{\mathbf{k}\boldsymbol{'}\mathbf{q}s}^{(2)}B_{\mathbf{kq}s}^{(1)}$. At 0~K,
the system can have equal numbers of $-(+)$ c-bosons, ``electrons''
(``hole'') composites, generated by $-v_0 B_{\mathbf{k}\boldsymbol{'}\mathbf{q}s}^{(1)\dagger}B_{\mathbf{kq}s}^{(2)\dagger}$.

The c-bosons, each with one fluxon, will be called the fundamental (f)
c-bosons. At a finite temperature, there are moving (non-condensed) fc
bosons. Their energies $w_q^{(j)}$ are obtained from \cite{Fujita_4}:
\begin{align}
w_q^{(j)} \Psi (\mathbf{k, q}) = &\varepsilon_{|\mathbf{k}\boldsymbol{+}\mathbf{q}|}^{(j)} \Psi(\mathbf{k, q}) - (2\pi\hbar)^{-2} v_0^* \nonumber\\
&\times \int' d^2 k' \Psi(\mathbf{k}\boldsymbol{'},\mathbf{q}),
\label{eq:eao_from}
\end{align}
where $\Psi(\mathbf{k, q})$ is the reduced wavefunction for the fc-boson; we neglected the
fluxon energy. The $v_0^*$ denotes the strength after the ladder diagram
binding, see below. For small $q$, we obtain a solution of Eq.~\eqref{eq:eao_from} as
\begin{equation}
w_q^{(j)} = \omega_0 + (2/\pi) v_\text{F}^{(j)} q, \quad w_0 = \frac{-\hbar \omega_\text{D}}{\exp(v_0^*D_0)^{-1}-1},
\label{eq:soE_as}
\end{equation}
where $v_\text{F}^{(j)}\equiv (2 \varepsilon_\text{F} / m_j)^{1/2}$ is
the Fermi velocity and $D_0 \equiv D(\varepsilon_\text{F})$ the density
of states per spin. The brief derivation of Eqs.\ \eqref{eq:eao_from} and \eqref{eq:soE_as} is
given in Appendix \ref{ap:Derivation_A}. Note that the energy $w_q^{(j)}$ depends
\emph{linearly} on the momentum magnitude $q$.

The system of free fc-bosons undergoes a Bose-Einstein condensation
(BEC) in 2D at the critical temperature \cite{Fujita_5}
\begin{equation}
k_\text{B} T_c = 1.24 \, \hbar v_\text{F} n_0^{1/2}.
\label{eq:critical_temperature}
\end{equation}
A brief derivation of Eq.~\eqref{eq:critical_temperature} is given in Appendix \ref{ap:Derivation_B}. The interboson
distance $R_0 \equiv n_0^{1/2}$ calculated from this expression is $1.24
\hbar v_\text{F} (k_\text{B}T_c)^{-1}$. The boson size $r_0$
calculated from Eq.~\eqref{eq:soE_as}, using the uncertainty relation ($q_\text{max}
r_0 \sim \hbar$) and $|w_0| \sim k_\text{B} T_c$ is $(2 / \pi) \hbar
v_\text{F} (k_\text{B}T_c)^{-1}$, which is a few times smaller than
$R_0$. Hence, the fc-bosons do not overlap in space, and the model of
free bosons is justified. For GaAs/AlGaAs, $m^*=0.067\,m_e$, $m_e$ =
electron mass. For the 2D electron density $10^{11}$\,cm$^{-2}$, we have
$v_\text{F} = 1.36\times 10^6$\,cm\,s$^{-1}$. Not all electrons are bound
with fluxons since the simultaneous generations of $\pm$ fc-bosons is
required. The minority carrier (``hole'') density controlls the fc-boson
density. For $n_0 = 10^{10}$\,cm$^{-2}$, $T_c = 1.29$\,K, which is
reasonable.

In the presence of Bose condensate below $T_c$ the unfluxed electron
carries the energy \cite{Fujita_5}
\begin{equation}
E_k^{(j)} = \sqrt{\varepsilon_k^{(f)2}+\triangle^2},
\label{eq:ect_energy}
\end{equation}
where the quasi-electron energy gap $\triangle$ is the solution of
\begin{align}
1 = &v_0 D_0 \int_0^{\hbar\omega_\text{D}} d\varepsilon \frac{1}{(\varepsilon^2 + \triangle^2)^{1/2}} \nonumber\\
&\times \left\{1 + \exp\left[-\beta(\varepsilon^2 + \triangle^2)^{1/2}\right]\right\}^{-1},\quad
\beta \equiv \frac{1}{k_\text{B}T}.
\label{eq:its_of}
\end{align}
Note that the gap $\triangle$ depends on the temperature $T$. At the
critical temperature $T_c$, there is no Bose condensate and hence $\triangle$
vanishes.

Now the moving fc-boson below $T_c$ has the energy $\tilde{w}_q$ obtained from
\begin{align}
\tilde{w}_q^{(j)}\Psi(\mathbf{k}, \mathbf{q}) = &E_{|\mathbf{k}\boldsymbol{+}\mathbf{q}|}^{(j)}
\Psi (\mathbf{k}, \mathbf{q}) - (2\pi\hbar)^{-2} v_0^* \nonumber\\
&\times \int' d^2 k' \Psi(\mathbf{k}\boldsymbol{'}, \mathbf{q}).
\label{eq:teo_from}
\end{align}
We obtain after solving Eq.~\eqref{eq:teo_from}:
\begin{equation}
\tilde{w}_q^{(j)} = \tilde{w}_0 + (2/\pi) v_\text{F}^{(j)} q \equiv w_0 +
\varepsilon_g + (2/\pi)v_\text{F}^{(j)}q,
\label{eq:oas_Eq}
\end{equation}
where $\tilde{w}_0(T)$ is determined from
\begin{equation}
1 = D_0\nu_0 \int_0^{\hbar \omega_\text{D}} d\varepsilon \left[\left|\tilde{w_0}\right| + (\varepsilon^2 + \triangle^2)^{1/2}\right]^{-1}.
\label{eq:wid_from}
\end{equation}
The energy difference:
\begin{equation}
\tilde{w}_0(T) - w_0 \equiv \varepsilon_g(T)
\label{eq:energy_difference}
\end{equation}
represents the $T$-dependent \emph{energy gap}. The energy $\tilde{w}_q$ is
negative. Otherwise, the fc-boson should break up. This limits $\varepsilon_g(T)$ to be
$|w_0|$ at 0~K. The fc-boson energy gap $\varepsilon_g$ declines to zero as the
temperature approaches $T_c$ from below.

The fc-boson, having the linear dispersion \eqref{eq:oas_Eq} can move in all
directions in the plane with the constant speed $(2 /
\pi)v_\text{F}^{(j)}$ as seen from Eq.~\eqref{eq:oas_Eq}. The supercurrent is
generated by the $\pm$ fc-bosons condensed monochromatically at the
momentum directed along the sample length. The supercurrent density
(magnitude) $J$, calculated by the rule: (charge $e^*$) $\times$
(carrier density $n_0$) $\times$ (drift velocity $v_d$), is
\begin{equation}
J \equiv e^* n_0 v_d = e^* n_0 (2 / \pi) \left| v_\text{F}^{(1)} - v_\text{F}^{(2)} \right|.
\label{eq:ddv_is}
\end{equation}
The induced Hall field (magnitude) $E_\text{H}$ equals $v_d B$. The
magnetic flux is quantized
\begin{equation}
BA = n_\phi (h / e), \quad n_\phi \text{ = fluxon density}.
\label{eq:mfi_quantized}
\end{equation}
Hence, we obtain
\begin{equation}
\rho_\text{H} \equiv \frac{E_\text{H}}{J} = \frac{v_d B}{e n_0 v_d} = \frac{1}{e
  n_0}n_\phi\left( \frac{h}{e} \right).
\label{eq:Hw_obtain}
\end{equation}
If $n_\phi = n_0$ valid at $\nu = 1$, we obtain $\rho_\text{H} = h / e^2$ in agreement with
the plateau value observed.

The model can be extended to the integer QHE at $\nu = P$, $P = 1, 2,
\dots$. The field magnitude is less. The LL degeneracy $eB\!A/h$ is
linear in $B$, and hence the lowest $P$ LL's must be considered. The
fc-boson density $n_0$ per LL is the electron density $n_e$ over $P$ and
the fluxon density $n_\phi$ is the boson density $n_0$ over $P$:
\begin{equation}
n_0 = n_e / P, \quad n_\phi = n_0 / P.
\label{eq:tbd_over}
\end{equation}
At $\nu = 1/2$ there are c-fermions, each with two fluxons. The c-fermions
have a Fermi energy. The $\pm$ c-fermions have effective masses. The
Hall resistivity $\rho_\text{H}$ has a $B$-linear behavior while the
resistivity $\rho$ is finite. In our theory the integer $P$ is the
number of the LL's occupied by the c-fermions.

Our Hamiltonian in Eq.~\eqref{eq:tpv_eliminated} can generate and stabilize the c-particles
with an arbitrary number of fluxons. For example, a c-fermion with two
fluxons is generated by two sets of the ladder diagram bindings, each
between the electron and the fluxon. The ladder diagram binding arises
as follows. Consider a hydrogen atom. The Hamiltonian contains kinetic
energies of the electron and the proton and the attractive Coulomb
interaction. If we regard the Coulomb interaction as a perturbation and
use a perturbation theory, we can represent the interaction process by
an infinite set of ladder diagrams, each ladder step connecting the
electron line and the proton line. The energy eigenvalues of this system is not
obtained by using the perturbation theory but they are obtained by
solving the Schr\"odinger equation directly. This example indicates that a
two-body bound state is represented by an infinite set of ladder
diagrams and that the binding energy (the negative of the ground-state
energy) is calculated by a non-perturbative method.

Jain introduced the effective magnetic field~\cite{Jain}
\begin{equation}
B^* \equiv B - B_\nu = B - (1/\nu) n_e (h/e)
\label{eq:effective_magnetic_field}
\end{equation}
relative to the standard field for the composite (c-)fermion. We extend
this idea to the bosonic (odd-denominator) fraction. This means that the
c-particle moves field-free at the exact fraction. The movement of the
guiding centers (the CM of the c-particle) can occur as if they are
subjected to no magnetic field at the exact fraction. The excess (or deficit)
of the magnetic field is simply the effective magnetic field $B^*$. The
plateau in $\rho_\text{H}$ is formed due to kind of the Meissner effect.
Consider the case of zero temperature near $\nu = 1$. Only the system
energy $E$ matters. The fc-bosons are condensed with the ground-state
energy $w_0$, and hence the system energy $E$ at $\nu = 1$ is $2N_0
w_0$, where $N_0$ is the number of $-$ fc-bosons (or $+$ fc-bosons). The
factor 2 arises since there are $\pm$ fc-bosons. Away from $\nu = 1$, we
must add the magnetic field energy $(2 \mu_0)^{-1}A(B^*)^2$, so that
\begin{equation}
E = 2N_0 w_0 + (2 \mu_0)^{-1}A(B^*)^2.
\label{eq:fes_that}
\end{equation}
When the field is reduced, the system tries to keep the same number
$N_0$ by sucking in the flux lines. Thus the magnetic field becomes
inhomogeneous outside the sample, generating the extra magnetic field
energy $(2\mu_0)^{-1}A(B^*)^2$. If the field is raised, the system tries
to keep the same number $N_0$ by expeling out the flux lines. The
inhomogeneous fields outside raise the field energy by $(2
\mu_0)^{-1}A(B^*)^2$. There is a critical field
$B_c^*=(4\mu_0|w_0|)^{1/2}$. Beyond this value, the superconducting
state is destroyed, which generates a symmetric exponential rise in the
resistance $R$. In our discussion of the Hall resistivity plateau we
used the fact that the ground-state energy $w_0$ of the fc-boson is
negative, that is, the c-boson is bound. Only then the critical field
$B_c^*=(4\mu|w_0|)^{1/2}$ can be defined. Here the phonon exchange
attraction played an important role. The repulsive Coulomb interaction,
which is the departure point of the prevalent fermionic theories~\cite{Ezawa,Jain},
cannot generate a bound state.

In the presence of the supercondensate, the non-condensed c-boson has an
energy gap $\varepsilon_g$. Hence, the non-condensed c-boson density has
the activation energy type exponential temperature-dependence:
\begin{equation}
\exp [-\varepsilon_g / (k_\text{B} T)].
\label{eq:ete_temperature-dependence}
\end{equation}
Some authors argue that the energy gap $\varepsilon_g$ for the integer
QHE is due to the LL separation = $\hbar\omega_0$. But the separation
$\hbar\omega_c$ is much greater than the observed $\varepsilon_g$.
Besides, from this view one cannot obtain the activation-type energy
dependence.

The BEC occurs at each LL, and therefore the c-boson density $n_0$ is
smaller for high $P$, see Eq.~\eqref{eq:tbd_over}, and the strengths become weaker as
$P$ increases. The most significant advantage of our bosonic theory is
that we are able to explain why the plateaus in the Hall resistivity
is developed when the resistivity is zero as the magnetic field is varied.
This plateau formation is phenomenologically assumed in the fermionic
theories. \cite{Ezawa,Jain}

\section{QUANTUM HALL EFFECT UNDER RADIATION}\label{se:QHE_Radiation}

The experiments by Mani \emph{et al}.\ \cite{Mani_1} and Zodov \emph{et
  al}.\ \cite{Zudov} indicate that the applied
radiation excites a large number of ``holes'' in the system. Using these
``holes'' and the preexisting ``electrons'' the phonon exchange can
pair-create $\pm$ c-bosons, that condense below $T_c$ in the excited
(upper) channel. The c-bosons condensed with the momentum along the
sample length are responsible for the supercurrent. In the presence of
the condensed c-bosons, the non-condensed c-bosons have an energy gap
$\varepsilon_g$, and therefore they are absent below $T_c$. The fermionic
currents in the base channel cannot be suppressed by the supercurrents
since the energy levels are different between the excited and base
channels. These c-fermions contribute a small normal current. They are subject to
the Lorentz force: $\mathbf{F} =
q(\mathbf{E}\boldsymbol{+}\mathbf{v}\boldsymbol{\times}\mathbf{B})$, and
hence they generate a Hall field $E_\text{H}$ proportional to the field
$B$. This is the main feature difference from the usual QHE (under no
radiation).

In the neighborhood of the QHE at $\nu = 1$, the current carriers in the base
and excited channels are, respectively, c-fermions and condensed
c-bosons. The currents are additive. We write down the total current
density $j$ as the sum of the fermionic current density $j_f$ and the
bosonic current density $j_b$:
\begin{equation}
j = j_f + j_b = en_f v_f + en_b v_b,
\label{eq:tbc_density}
\end{equation}
where $v_f$ and $v_b$ are the drift velocities of the c-fermions and
c-bosons. The Hall fields $E_\text{H}$ are additive, too. Hence we have
\begin{equation}
E_\text{H} = E_{\text{H},f} + E_{\text{H},b} = v_f B + v_b B.
\label{eq:tHw_have}
\end{equation}
The Hall effect condition ($E_\text{H} = v_d B$) applies separately
for the c-fermions and c-bosons. We therefore obtain
\begin{equation}
R_\text{H} = \frac{E_\text{H}}{j} = \frac{v_f B + v_b B}{n_f v_f + n_b v_b}\frac{1}{e}.
\label{eq:cWt_obtain}
\end{equation}
Far away from the midpoint of the zero-resistance stretch, the c-bosons
are absent and hence the Hall resistivity $R_\text{H}$ becomes $B/(en_f)$:
\begin{equation}
R_\text{H} = B/(en_f)\quad \text{(far away)},
\label{eq:tHr_becomes}
\end{equation}
after the cancellation of $v_f$. At the midpoint the c-bosons are
dominant. Then, the Hall resistivity $R_\text{H}$ is approximately
equal to $h/e^2$ since
\begin{equation}
\frac{E_\text{H}}{j} \cong \frac{v_b B}{e n_b v_b} \cong
\frac{h}{e^2}\frac{n_\phi}{n_b} = \frac{h}{e^2}\quad \text{(midpoint)},
\label{eq:aet_since}
\end{equation}
where we used the flux quantization [$B = (h/e)n_\phi$], and the fact
that the flux density $n_\phi$ equals the c-boson density $n_b$ at $\nu=1$. The
Hall resistivity $R_\text{H} = E_\text{H} / j$ is not exactly equal to
$h / e^2$ since the c-fermion current density $en_f v_f$ is much smaller
than the supercurrent density $en_b v_b$, but it does not vanish. In the
horizontal stretch the system is superconducting, and hence the
supercurrent dominates the normal current: $e n_b v_b \gg e n_f v_f$.
The deviation $\Delta R_\text{H}$ is, using Eq.~\eqref{eq:cWt_obtain},
\begin{align}
\Delta R_\text{H} &= \frac{v_f B + v_b B}{e(n_f v_f + n_b v_b)} -
\frac{v_f B}{e n_f v_f} \nonumber\\
&\simeq \frac{n_b B}{e n_b v_b} \simeq \frac{h}{e^2}.
\label{eq:diu_Eq}
\end{align}
If the field $B$ is raised (or lowered) a little from the midpoint,
$\Delta R_\text{H}$ is a constant ($h/e^2$) due to the Meissner effect.
If the field is raised high enough, the superconducting state is
destroyed and the normal current sets in, generating a finite resistance
and a vanishing $\Delta R_\text{H}$. Hence the deviation $\Delta
R_\text{H}$ and the diagonal resistance $R_{xx}$ are closely correlated
as observed by Mani~\cite{Mani_1}.

In Fig.~2 of Ref.~\cite{Mani_1} (not shown here), we can see that in the range,
where the SdH oscillations are observed for the
resistance without radiation, the signature of oscillations also appear
for the resistance $R_{xx}$ with radiation. The SdH oscillations arise
only for the fermion carriers. This SdH signature
in $R_{xx}$ should remain. Our two-channel model is supported here.

Mani \emph{et al}., Fig.~2 of Ref.~\cite{Mani_1}, shows that the
strength of the superconducting state does not change much. The 2D
density of states for the conduction electrons associated with the
circular Fermi surface is independent of the electron energy, and hence
the number of the excited electrons is roughly independent of the
radiaton energy (frequency). The ``hole''-like excitations are absent
with no radiation. We suspect that the ``hole''-band edge is a distance
$\varepsilon_0$ away from the system's Fermi level. This means that if
the radiation energy $\hbar \omega$ is less than $\varepsilon_0$, the
radiation can generate no superconducting state. An experimental
confirmation is highly desirable here.

If a bias voltage is applied to the system, then a normal current runs
in the base channel. In the upper channel the supercurrent still runs
with no potential drop. The both currents run in the same sample space. The
apparent discrepancy in the electric potential here may be resolved by
considering a static charge $Q$ developed in the system upon the field application.
That is, the system will be charged and the static potential
\begin{equation}
V_c = \frac{1}{2} C Q^2,
\label{eq:static_potential}
\end{equation}
where $C$ is the system's capacitance, can balance the total electric
potential while the charging does not affect the supercurrents. This
effect may be checked by experiments, which is a critical test for our
two-channel model.

In summary, the QHE under radiation is the QHE at the upper channel. The
condensed c-bosons generate a superconducting state with a gap
$\varepsilon_g$ in the c-boson energy spectrum. The supercondensate
suppresses the c-particle currents in the upper channel, but cannot suppress
the normal currents in the base channel. Thus, there is a finite
resistive current accompanied by the Hall field. This explains the
$B$-linear Hall resistivity.

\section{THE DENSITY OF STATES AND STATISTICAL WEIGHT}\label{se:DOS_SW}

We calculate magnetic oscillations in the statistical weight for a 2D
electron system. Let us take a dilute system of electrons moving in the
plane. Applying a magnetic field $\mathbf{B}$ perpendicular to the
plane, each electron will be in the Landau states with the energy given
by
\begin{equation}
E = (N_\text{L}+1/2) \hbar \omega_c,\quad N_\text{L} = 0,1,2,\dots.
\label{eq:teg_by}
\end{equation}
The degeneracy of the Landau level (LL) is
\begin{equation}
e B A / 2 \pi \hbar, \quad A = \text{sample area}.
\label{eq:LlL_is}
\end{equation}
The weaker the field the more LL's, separated by $\hbar \omega_c$, are
occupied by the electrons. The electron in the Landau state can be viewed as
circulating around the guiding center.

\begin{figure}
\centering
\includegraphics{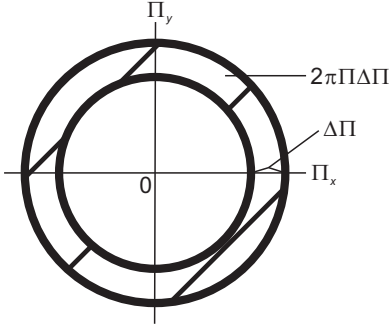}
\caption{A 2D Landau state is represented by the ring (shaded area) of
  the phase-space volume $2\pi \Pi \Delta \Pi$.}\label{fg:Figure1}
\end{figure}

We introduce kinetic momenta
\begin{equation}
\Pi_x = p_x + eA_x,\quad \Pi_y = p_y + eA_y,
\label{eq:kinetic_momenta}
\end{equation}
in terms of which the Hamiltonian $\mathcal{H}$ for the electron is
\begin{equation}
\mathcal{H} = \frac{1}{2m^*}\left(\Pi_x^2 + \Pi_y^2\right) \equiv \frac{1}{2m^*}\Pi^2.
\label{eq:fte_is}
\end{equation}
The vector potential $\mathbf{A} = (1/2) \mathbf{B} \boldsymbol{\times} \mathbf{r}$ can be written as $A_x =
-(1/2)By$, $A_y = (1/2)Bx$, $A_z = 0$. Using the quantum condition $[x,
  p_x] = [y, p_y] = i \hbar$, $[x, y] = [p_x, p_y] = 0$, we obtain
\begin{equation}
[\Pi_x, \Pi_y] = -(e \hbar / i) B.
\label{eq:qcw_obtain}
\end{equation}
If we introduce
\begin{equation}
(m^*)^{1/2} \Pi_x \equiv P,\quad (eB)^{-1}(m^*)^{1/2}\Pi_y \equiv Q,
\label{eq:Iw_introduce}
\end{equation}
we obtain
\begin{equation}
\mathcal{H} = (1/2)\left[ P^2 + \omega_c^2Q^2 \right],
\label{eq:w_obtain}
\end{equation}
and
\begin{equation}
[Q, P] = i\hbar.
\label{eq:and}
\end{equation}
Hence, the energy eigenvalues are given by $(N_\text{L} + 1/2) \hbar \omega_c$,
confirming Eq.~\eqref{eq:teg_by}. After simple calculations, we obtain
\begin{equation}
dx d\Pi_x dy d\Pi_y = dx dp_x dy dp_y.
\label{eq:scw_obtain}
\end{equation}
We can then represent quantum states by small quasi-phase space cell of
the volume $dxd\Pi_xdyd\Pi_y$. The Hamiltonian $\mathcal{H}$ in Eq.~\eqref{eq:fte_is} does not
depend on the position $(x, y)$. Assuming large normalization lengths
$(L_1, L_2)$, we can represent the Landau states by the concentric shells
of the phase space having the statistical weight
\begin{equation}
2 \pi \Pi \Delta \Pi \cdot L_1 L_2(2\pi\hbar)^{-2} =
\frac{eBA}{2\pi\hbar},
\label{eq:statistical_weight}
\end{equation}
where A = $L_1 L_2$ and $\hbar \omega_c = \Delta (\Pi^2 / 2m^*) = \Pi
\Delta \Pi / m^*$. Hence, the LL degeneracy is given by Eq.~\eqref{eq:LlL_is}.
Figure~\ref{fg:Figure1} represents a typical Landau state in the
$\Pi_x$-$\Pi_y$ space.%

As the field $B$ is raised the separation $\hbar\omega_c$ increases, and
the quantum states are bunched together. As a result of the bunching,
the density of states $\mathcal{N}(\varepsilon)$ should change periodically.

The electrons obey the Fermi-Dirac statistics. Considering a system of
free electrons, we define the Helmholtz free energy $\mathcal{F}$ by
\begin{equation}
\mathcal{F} = N \mu - 2 k_\text{B} T \sum \ln \left[1 + e^{(\mu - E_i)/k_\text{B} T} \right],
\label{eq:Helmholtz_free_energy}
\end{equation}
where $\mu$ is the chemical potential and the factor 2 arises from the
spin degeneracy. The chemical potential $\mu$ is determined from the
condition
\begin{equation}
\frac{\partial \mathcal{F}}{\partial \mu} = 0.
\label{eq:dft_condition}
\end{equation}
The total magnetic moment $M$ for the system can be found from
\begin{equation}
M = -\frac{\partial \mathcal{F}}{\partial B}.
\label{eq:cbf_from}
\end{equation}
Equation \eqref{eq:cbf_from} is equivalent to the usual condition that the total
number of the electrons, $N$, can be obtained in terms of the Fermi
distribution function
\begin{equation}
N = 2 \sum_i f(E_i).
\label{eq:Fermi_distribution_function}
\end{equation}
The LL $E_i$ is characterized by the Landau oscillator quantum number $N_\text{L}$.
Let us introduce the \emph{density of state}
\begin{equation}
dW / dE \equiv \mathcal{N}(E)
\label{eq:density_of_state}
\end{equation}
such that $\mathcal{N}(E) dE$ = the number of states having an energy
between $E$ and $E + dE$. We write Eq.~\eqref{eq:Helmholtz_free_energy} in the form
\begin{align}
\mathcal{F} &= N \mu - 2 k_\text{B} T \int_0^\infty dE \frac{dW}{dE}\ln \left[1 +
  e^{(\mu - E)/k_\text{B} T} \right] \nonumber\\
  &= N \mu - 2 k_\text{B} T \int_0^\infty dE W(E) f(E).
\label{eq:Eit_form}
\end{align}

The statistical weight (number) $W$ is the total number of states having
energies less than
\begin{equation}
E = (N_\text{L} + 1/2) \hbar \omega_c.
\label{eq:hel_than}
\end{equation}
For a fixed pair $(E, N_\text{L})$, the density of states is
\begin{equation}
dW = \frac{L_1L_2}{(2\pi \hbar)^2}2\pi \Pi \Delta\Pi \Theta [E - (N_\text{L} +
1/2) \hbar \omega_c],
\label{eq:dos_is}
\end{equation}
where $\Theta(x)$ is the Heaviside step function:
\begin{equation}
\Theta(x) = \left\{
\begin{array}{cl}
1 & \text{if } x > 0 \\
0 & \text{if } x < 0
\end{array}\right..
\label{eq:Heaviside_step_function}
\end{equation}
We sum Eq.~\eqref{eq:dos_is} with respect to $N_\text{L}$ and obtain
\begin{equation}
W(E) = C(\hbar \omega_c)2 \sum_{N_\text{L} = 0}^\infty \Theta [\varepsilon
- (2N_\text{L} + 1) \pi],
\label{eq:rta_obtain}
\end{equation}
\begin{equation}
C = 2\pi m^* A (2 \pi \hbar)^{-2},\quad \varepsilon \equiv 2\pi E
/ \hbar \omega_c.
\label{eq:rta_obtain_2}
\end{equation}

We assume a high Fermi-degeneracy such that
\begin{equation}
\mu \simeq \varepsilon_\text{F} \gg \hbar \omega_c.
\label{eq:high_Fermi-degeneracy}
\end{equation}
The sum in Eq.~\eqref{eq:rta_obtain} can be computed by using Poisson's summation
formula \cite{Courant}
\begin{equation}
\sum_{n=-\infty}^{\infty} f(2\pi n) = \frac{1}{2\pi} \sum_{m =
-\infty}^{\infty} \int_{-\infty}^{\infty}d\tau f(\tau) e^{-i \omega \tau}.
\label{eq:Poissons_summation_formula}
\end{equation}
We then obtain \cite{Mani_2}
\begin{align}
& W(E) = W_0 + W_\text{osc},\label{eq:Wt_obtain} \\
& W_0 = A (m^* / \pi \hbar^2) E\label{eq:Wt_obtain_2} \\
& W_\text{osc} =
C \hbar \omega_c \frac{2}{\pi} \sum_{\nu = 1}^\infty \frac{(-1)^\nu}{\nu} \sin \left( \frac{2\pi\nu E}{\hbar \omega_c} \right).
\label{eq:Wt_obtain_3}
\end{align}
The detailed calculations leading to Eqs.\ \eqref{eq:Wt_obtain}--\eqref{eq:Wt_obtain_3} are given in Appendix
\ref{ap:SW_LS}. Only the first term $\nu = 1$ in Eq.~\eqref{eq:Wt_obtain_3} will be important in
practice for weak fields
\begin{equation}
\varepsilon_\text{F} \gg \hbar\omega_c,
\label{eq:pfw_fields}
\end{equation}
which will be shown later.

The term $W_0$, which is independent of $B$, gives the weight equal to
that for a free electron system with no field. Note that there are no
Landau-like diamagnetic terms proportional to the squared field $B^2$.

\section{SHUBNIKOV-DE HAAS OSCILLATIONS}\label{se:SdH_Oscillations}

Let us, first, consider the case with no magnetic field. We assume a
uniform distribution of impurities with density $n_I$. We introduce a
momentum distribution function $\phi (\mathbf{p},t)$, defined such that
$\phi (\mathbf{p},t) d^2p$ gives the relative probability of finding an
electron in the element $d^2p$ at time $t$. This function will be
normalized such that
\begin{equation}
\frac{2}{(2\pi \hbar)^2}\int d^2 p \, \phi(\mathbf{p},t) = \frac{N}{A} \equiv n,
\label{eq:bns_that}
\end{equation}
where the factor 2 is due to the spin degeneracy.

The electric current density $\mathbf{j}$ is given in terms of
$\phi(\mathbf{p},t)$ as
\begin{equation}
\mathbf{j} = \frac{-2e}{(2\pi\hbar)^2 m^*} \int d^2p \, \phi(\mathbf{p},t) \mathbf{p}.
\label{eq:ito_as}
\end{equation}
The distribution function $\phi (\mathbf{p},t)$ can be obtained by
solving the Boltzmann equation for the stationary homogeneous state,
dropping $t$:
\begin{equation}
e \mathbf{E} \boldsymbol{\cdot} \frac{\partial}{\partial\mathbf{p}} \phi(\mathbf{p})
= \frac{n_I}{m^*} \int d\Omega\,
I(p,\theta) [\phi(\mathbf{p}\boldsymbol{'}) - \phi(\mathbf{p})]p,
\label{eq:shs_dropping}
\end{equation}
where $\theta$ is the scattering angle, that is, the angle between the initial
momentum $\mathbf{p}$ and the final one $\mathbf{p}\boldsymbol{'}$, and
$I(p,\theta)$ is the differential cross section. Solving Eq.~\eqref{eq:shs_dropping}, we
obtain the conductivity as \cite{Fujita_3}
\begin{equation}
\sigma = \frac{2}{(2\pi\hbar)^2}\frac{e^2}{m^*}\int
d^2p\left(-\frac{df}{dE}\right)\frac{E}{\Gamma(E)},\quad E \equiv \frac{p^2}{2m^*},
\label{eq:otc_as}
\end{equation}
where $\Gamma$ is the energy ($E$)-dependent relaxation rate
\begin{equation}
\Gamma (E) = n_I \int d\Omega I(p,\theta)(1 - \cos\theta)\frac{p}{m^*}.
\label{eq:E-dependent_relaxation_rate}
\end{equation}
The Fermi distribution function
\begin{equation}
f(E) \equiv \left[e^{\beta (E-\mu)} + 1\right]^{-1}
\label{eq:Fermi_distribution_function_2}
\end{equation}
is normalized such that
\begin{align}
n &= \frac{2}{(2\pi\hbar)^2}\int d^2 p f(E) \nonumber\\
&= \int_0^\infty dE \nu(E)f(E),\quad \nu(E) \equiv \frac{\mathcal{N}(E)}{A},
\label{eq:ins_that}
\end{align}
where $\nu(E)$ is the density of states per area. We can rewrite
Eq.~\eqref{eq:otc_as} as
\begin{equation}
\sigma = \frac{e^2}{m^*} \int_0^\infty dE \nu(E) \left( -\frac{df}{dE}
\right) \frac{E}{\Gamma(E)}.
\label{eq:crE_as}
\end{equation}

The Fermi distribution function $f(E)$ drops steeply near $E = \mu$ at
low temperatures: $k_\text{B} T \ll \varepsilon_\text{F}$ (Fermi energy).
If the density of states varies slowly with energy $E$, then the
delta-function replacement formula
\begin{equation}
-\frac{df}{dE} = \delta(E - \mu)
\label{eq:delta-function_replacement_formula}
\end{equation}
can be used. Using
\begin{equation}
\int_0^\infty dE \mathcal{N}(E) \left(-\frac{df}{dE}\right) E =
\int_0^\infty dE \mathcal{N}(E) f(E),
\label{eq:cbu_Using}
\end{equation}
and comparing Eqs.\ \eqref{eq:crE_as} and the Drude formula
\begin{equation}
\sigma = \frac{e^2}{m^*} n \frac{1}{\gamma_0},
\label{eq:Drude_formula}
\end{equation}
we obtain
\begin{equation}
\frac{n}{\gamma_0(T)} = \int_0^\infty dE \nu(E) f(E)\frac{1}{\Gamma_0(E)}.
\label{eq:w_obtain_2}
\end{equation}
Note that the temperature dependence of the relaxation rate $\gamma_0(T)$ is
introduced through the Fermi distribution function $f(E)$.

Let us now consider a field-dressed electron (guiding center). We assume
that the dressed electron is a fermion with magnetotransport mass $M^*$
and charge $e$. The kinetic energy is represented by
\begin{equation}
\mathcal{H'} = \frac{1}{2M^*} \left(\Pi_x^2 + \Pi_y^2\right) \equiv \frac{1}{2M^*}\Pi^2.
\label{eq:eir_by}
\end{equation}

We introduce a \emph{distribution function} $\varphi(\mathbf{\Pi}, t)$ in the
$\Pi_x \Pi_y$-space normalized such that
\begin{equation}
\frac{2}{(2 \pi \hbar)^2} \int d^2 \Pi \,\varphi (\Pi_x, \Pi_y, t)
= \frac{N}{A} = n.
\label{eq:sns_that}
\end{equation}
The Boltzmann equation for a homogeneous stationary state of the system
is
\begin{equation}
e (\mathbf{E} + \mathbf{v} \boldsymbol{\times} \mathbf{B}) \boldsymbol{\cdot} \frac{\partial \varphi}{\partial \mathbf{\Pi}} = \int
d\Omega \frac{\Pi}{M^*} n_I I(\Pi, \theta) [\varphi(\mathbf{\Pi}\boldsymbol{'}) - \varphi(\mathbf{\Pi})],
\label{eq:ots_is}
\end{equation}
where $\theta$ is the scattering angle, that is, the angle between the
initial and final kinetic momenta ($\mathbf{\Pi}, \mathbf{\Pi}\boldsymbol{'}$). In the actual
experimental condition, the magnetic force term can be neglected.
Assuming this condition, we obtain the same Boltzmann equation \eqref{eq:shs_dropping} for a
field-free system except the mass difference. Hence, we obtain
\begin{equation}
\sigma = \frac{2 e^2}{M^*(2\pi\hbar)^2}\int d^2
p \frac{E}{\Gamma}\left( -\frac{d f}{d E} \right).
\label{eq:dHw_obtain}
\end{equation}

As the field $B$ is raised, the separation $\hbar \omega_c$ becomes
greater and the quantum states are bunched together. The statistical
weight $W$ contains an oscillatory part, see Eq.~\eqref{eq:Wt_obtain_3}
\begin{equation}
W_\text{osc} \propto \sin \left( \frac{2\pi\varepsilon'}{\hbar\omega_c} \right),\quad \varepsilon' = \frac{\Pi^{'2}}{2m^*}.
\label{eq:ops_Eq}
\end{equation}
Physically, the sinusoidal variations in Eq.~\eqref{eq:Wt_obtain_3} arise as follows. From the Heisenberg
uncertainty principle (phase space consideration) and the Pauli
exclusion principle, the Fermi energy $\varepsilon_\text{F}$ remains
approximately unchanged as the field $B$ varies. The density of states is
high when $\varepsilon_\text{F}$ matches the $N_\text{L}$-th level,
while it is small when $\varepsilon_\text{F}$ falls between neighboring LLs.

If the density of states, $\mathcal{N}(\varepsilon)$, oscillates
violently in the drop of the Fermi distribution function
$f(\varepsilon) \equiv \left[e^{\beta (\varepsilon - \mu)} + 1\right]^{-1}$, one
cannot use the delta-function replacement formula \eqref{eq:delta-function_replacement_formula}. The width of
$d f / d \varepsilon$ is of the order $k_\text{B}T$. The critical
temperature $T_c$ below which the oscillations can be observed is
\begin{equation}
k_\text{B} T_c \sim \hbar \omega_c.
\label{eq:cbo_is}
\end{equation}
Below $T_c$, we may proceed as follows. Let us consider the integral
\begin{equation}
I = \int_{0}^{\infty} d E
f(E) \sin \left( \frac{2\pi E}{\hbar \omega_c} \right), \quad E \equiv \frac{\Pi^2}{2
M^*}.
\label{eq:uct_integral}
\end{equation}
We introduce a new variable $\zeta \equiv \beta(E - \nu)$, and extend the
lower limit to $-\infty$ (low temperature limit):
\begin{align}
\int_0^\infty d E \cdots \frac{1}{e^{\beta(E-\mu)}+1} &= \frac{1}{\beta}\int_{-\mu\beta}^\infty
d \zeta \cdots \frac{1}{e^\zeta + 1} \nonumber\\
&\rightarrow \frac{1}{\beta} \int_{-\infty}^\infty d\zeta\cdots \frac{1}{e^\zeta + 1}.
\label{eq:tlt_limit}
\end{align}
With the help of the integral formula
\begin{equation}
\int_{-\infty}^\infty d \zeta \frac{e^{\text{i}\alpha \zeta}}{e^\zeta + 1} =
\frac{\pi}{\text{i}\sinh \pi\alpha},
\label{eq:integral_formula}
\end{equation}
which is proved in Appendix \ref{ap:Derivation_C}, we obtain from Eq.~\eqref{eq:uct_integral}:
\begin{equation}
I = -\frac{\pi k_\text{B} T \cos(2 \pi \varepsilon_\text{F}
/ \hbar \omega_c)}{\sinh (2 \pi^2 M^* k_\text{B} T / \hbar e B)}.
\label{eq:wof_Eq}
\end{equation}
Here, we used
\begin{equation}
M^* \mu (T = 0) = m^* \varepsilon_\text{F},
\label{eq:Hw_used}
\end{equation}
since the Fermi momentum is the same for both dressed and undressed
electrons. For very low fields, the oscillation number in the range
$k_\text{B}T$ becomes great, and hence the sinusoidal contribution must
cancel out. This effect is represented by the factor $[\sinh (2\pi^2 M^* k_\text{B} T / \hbar e
 B)]^{-1}$.

We now calculate the conductivity, starting with Eq.~\eqref{eq:crE_as}. For the
field-free case, we may use Eqs.\ \eqref{eq:delta-function_replacement_formula} and \eqref{eq:cbu_Using} to obtain
\begin{equation}
\frac{n}{\gamma_0}
= \frac{\nu(\varepsilon_\text{F})\varepsilon_\text{F}}{\Gamma_0(\varepsilon_\text{F})}.
\label{eq:Eat_obtain}
\end{equation}
For a finite $B$, the non-oscillatory part (background) contributes a
similar amount:
\begin{equation}
\frac{n}{\gamma}
= \frac{\nu(\varepsilon_\text{F})\varepsilon_\text{F}}{\Gamma(\varepsilon_\text{F})},
\label{eq:cas_amount}
\end{equation}
calculated for the dressed electrons. The oscillatory part can be
calculated by using the integration formula $I$ in Eqs.\ \eqref{eq:uct_integral} and
\eqref{eq:wof_Eq}. This part is much smaller than
$\nu(\varepsilon_\text{F})\varepsilon_\text{F} / \Gamma(\varepsilon_\text{F})$
in Eq.~\eqref{eq:cas_amount}, since the contribution is limited to the small energy
range $k_\text{B} T$. It is also small by the sinusoidal cancellation.
We, therefore, obtain
\begin{align}
\frac{n}{\gamma} &=
\frac{\nu(\varepsilon_\text{F})\varepsilon_\text{F}}{\Gamma(\varepsilon_\text{F})}(1 + \phi),\label{eq:cWt_obtain_2} \\
\phi &\equiv \frac{\pi k_\text{B}
T}{\varepsilon_\text{F}} \frac{\cos(2 \pi \varepsilon_\text{F}
/ \hbar \omega_c)}{\sinh (2 \pi^2 M^*k_\text{B} T / \hbar eB)}.
\label{eq:cWt_obtain_3}
\end{align}

Strictly speaking, the contribution of the terms with $\nu = 2,3,\cdots$
in the sum $W_\text{osc}$ in Eq.~\eqref{eq:Wt_obtain_3} should be added. But this
contribution, which carries $[\sinh (2\pi^2 \nu M^* k_\text{B} T / \hbar e B)]^{-1}$,
is small since
\begin{equation}
\sinh (2\pi^2 M^* k_\text{B} T / \hbar e B) \gg 1.
\label{eq:cis_since}
\end{equation}
In the present theory, the two masses $m^*$ and $M^*$ are introduced
naturally corresponding to the two physical processes: the cyclotron
motion of the electron and the guiding center motion of the dressed
electron. The dressed electron is the same entity as the c-fermion with
two fluxons in the QHE theory.

In summary, the magnetoconductivity $\sigma(B)$, given by Eq.~\eqref{eq:dHw_obtain}, may be written out as
\begin{equation}
\sigma = \frac{e^2}{M^*}\frac{n}{\gamma} = \frac{e^2}{M^*}
\frac{\nu (\varepsilon_\text{F})\varepsilon_\text{F}}{\Gamma(\varepsilon_\text{F})}(1 + \phi).
\label{eq:bwo_as}
\end{equation}
In contrast, the conductivity $\sigma_0$ at zero field is
\begin{equation}
\sigma_0 = \frac{e^2}{m^*}\frac{n}{\gamma_0} = \frac{e^2}{m^*}
\frac{\nu(\varepsilon_\text{F})\varepsilon_\text{F}}{\Gamma_0(\varepsilon_\text{F})},
\label{eq:azf_is}
\end{equation}
where we have assumed that the Fermi energy $\varepsilon_\text{F}$
remains the same for both cases. We note that the magnetoconductivity
$\sigma$ does not approach the conductivity $\sigma_0$ in the low field
limit. In fact, we obtain in this limit ($\phi=0$):
\begin{equation}
\sigma - \sigma_0 = e^2 n \left(\frac{1}{M^*\gamma} - \frac{1}{m^*\gamma_0} \right).
\label{eq:oit_limit}
\end{equation}
The difference arises from the carrier difference.

If the ``decay'' rate $\delta = 2 \pi^2 M^*k_\text{B} T / \hbar e$ defined through
\begin{equation}
\sinh (\delta / B) \equiv \sinh (2 \pi^2 M^*k_\text{B} T / \hbar eB)
\label{eq:drd_through}
\end{equation}
is measured carefully, the magnetotransport mass $M^*$ can be obtained
\emph{directly} through
\begin{equation}
M^* = e\hbar\delta / (2 \pi^2 k_\text{B} T).
\label{eq:bod_through}
\end{equation}

\begin{figure}
\centering
\includegraphics{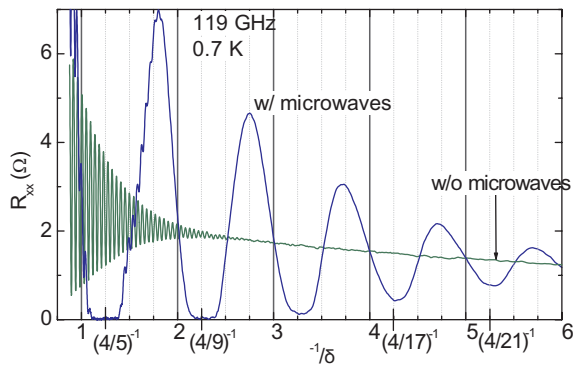}
\caption{The resistance $R_{xx}$ versus the reduced inverse magnetic
  field, $B'^{-1}$. See Mani~\cite{Mani_2} for the actual reduction. The number
  $N$ in the abscissas is the intersection number between the curves
  with (w/) and without (w/o) microwaves.}\label{fg:Figure2}
\end{figure}

Mani measured the SdH oscillations in GaAs/AlGaAs \cite{Mani_2}, Fig.~1,
$T=0.7$\,K. His data are reproduced in Fig.~\ref{fg:Figure2}.%
Clearly, we see the diagonal resistance $R_{xx}$ linearly decreasing
with $B^{-1}$ in the low field limit. For high purity samples at very
low temperature ($\sim 0.7$\,K), the impurity and phonon scatterings are
negligible. By the energy-time uncertainty principle the dressed
electron can spend a short time at the upper LL and come back to the
ground LL with a different guiding center, which causes a guiding center
jump. We assume that the relaxation rate $\gamma$ is the natural
linewidth arising from the LL separation divided by $\hbar$, that is,
the cyclotron frequency $\omega_c$:
\begin{equation}
\gamma = \omega_c = e B / m^*.
\label{eq:cyclotron_frequency}
\end{equation}
This generates the desired $B^{-1}$ dependence for $R_{xx}$.

We fitted Mani's data in Fig.~\ref{fg:Figure2} with
\begin{equation}
R_{xx} = A + Bx + \frac{[E \cos(2\pi Cx) + F] x}{\sinh (Dx)},
\label{eq:diF_with}
\end{equation}
where $A = 2.3$, $B = -0.18$, $C = 23.0$, $D = 3.1$, $E = 22.0$, and $F =
7.0$. The fits agree with the data within the experimental errors. Using
$Dx = \delta / B$, we obtain
\begin{equation}
M^* = 0.30\,m_e,
\label{eUw_obtain}
\end{equation}
where $m_e$ is the gravitational electron mass. If $m^*=0.067\,m_e$, then
$M^* / m^* = 4.5$. These are reasonable numbers.

The relaxation rate $\gamma = \Gamma(\varepsilon_\text{F})$ can now be
obtained through Eq.~\eqref{eq:Drude_formula} with the measured magnetoconductivity. All
electrons, not just those excited electrons near the Fermi surface, are
subject to the electric field. Hence, the carrier density $n$ appearing
in Eq.~\eqref{eq:bwo_as} is the total density $n$ of the dressed electrons. This
$n$ also appears in the Hall resistivity expression
\begin{equation}
\rho_\text{H} \equiv \frac{E_\text{H}}{j} = \frac{v_d B}{e n v_d} = \frac{B}{en},
\label{Hall_resistivity}
\end{equation}
where the Hall effect condition:
\begin{equation}
E_\text{H} = v_d B, \quad v_d = \text{drift velocity}
\label{eq:Hall_effect_condition}
\end{equation}
was used.

The dressed electrons are there whether the system is probed in
equilibrium or in nonequilibrium as long as the system is subjected to a
magnetic field. Hence their presence can be checked by measuring the
susceptibility or the heat capacity of the system. All (dressed)
electrons are subject to the magnetic field, and hence the magnetic
susceptibility $\chi$ is proportional to the carrier density $n$
although the $\chi$ depends critically on the Fermi surface. We shall
briefly discuss the magnetic moment and susceptibility.

The magnetization $\mathcal{M}$, that is, the total magnetic moment per
unit area, can be obtained from
\begin{equation}
\mathcal{M} = -\frac{\partial \mathcal{F}}{\partial B}.
\label{eq:cbo_from}
\end{equation}
Using Eqs.\ \eqref{eq:dft_condition} and \eqref{eq:high_Fermi-degeneracy}, we obtain the magnetization $\mathcal{M}$ for the
quasi-free electrons \cite{Fujita_3}
\begin{multline}
\mathcal{M} = 2n \frac{\mu_\text{B}^2}{\varepsilon_\text{F}}
\bigg[1 - \left( \frac{\varepsilon_\text{F}}{\mu_\text{B}B} \right)
\frac{k_\text{B}T}{\varepsilon_\text{F}} \left( \frac{m^*}{M^*} \right)\\
\times \frac{\cos (2\pi \varepsilon_\text{F} / \hbar
 \omega_c)}{\sinh (2 \pi^2 M k_\text{B} T / \hbar eB)}\bigg],
\label{eq:ftq_electrons}
\end{multline}
where $\mu_\text{B}$ is the Bohr magneton. The magnetic susceptibility
$\chi$ is defined by the ratio
\begin{equation}
\chi \equiv \frac{\mathcal{M}}{B}.
\label{eq:dbt_ratio}
\end{equation}

\section{TRANSPORT DIAMOND AND ZERO CURRENT ANOMALY}\label{se:TD_ZCA}

\begin{figure}
\centering
\includegraphics{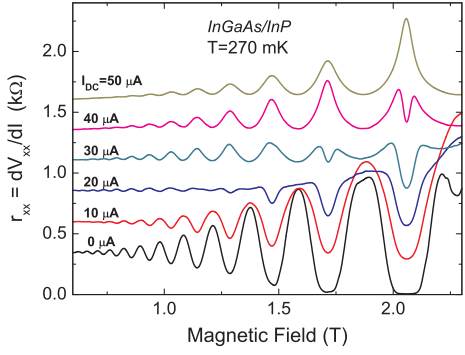}
\caption{The differential resistance $r$ of an InGaAs/InP Hall bar
  (width = $100\,\mu$m) at different DC values, $T=270$\,mK. All
  curves except at $I_\text{DC}=0$ are shifted vertically by
  $0.25$\,k$\Omega$ for clarity.}\label{fg:Figure3}
\end{figure}

We are now ready to discuss the TD and ZCA observed by Studenikin
\emph{et al}.\ \cite{Studenikin}. Ref.~\cite{Studenikin}, Fig.~1 is reproduced in Fig.~\ref{fg:Figure3}.%
The outstanding features are:
\begin{enumerate}
\item[(A)] The differential resistance $r \equiv dV / dI$ exhibit the SdH
oscillations for higher DC, $I_\text{DC} = 50\,\mu$A. The
envelope of the SdH oscillations become smaller for weaker magnetic
fields.

\item[(B)] The background differential resistance for the SdH is zero.

\item[(C)] The flat minima present at $I_\text{DC} = 0$ indicate a QHE. The
flat minimum means a zero resistance
\begin{equation}
R \equiv V/I = 0.
\label{eq:maz_resistance}
\end{equation}

\item[(D)] The SdH maxima and the QHE minima both have the right-left symmetry
with varying magnetic fields.

\item[(E)] As the DC increases, the SdH maxima progressively become the QHE minima.
\end{enumerate}
Our interpretation is as follows.
\begin{enumerate}
\item[(A)] The SdH oscillations are described by formula \eqref{eq:cWt_obtain_3}. The
oscillations are sinusoidal:
\begin{equation}
\cos (2 \pi \varepsilon_\text{F} / \hbar \omega_c) =
\cos (2 \pi m^* \varepsilon_\text{F} / \hbar e B),
\label{eq:Toa_sinusoidal}
\end{equation}
and the envelope is represented by
\begin{equation}
\frac{\pi k_\text{B} T}{\varepsilon_\text{F}}\frac{1}{\sinh (2 \pi^2 M^*
k_\text{B} T /\hbar e B)}.
\label{eq:eir_by_2}
\end{equation}
The cyclotron mass $m^*$ appears in Eq.~\eqref{eq:Toa_sinusoidal} and the magnetotransport mass
$M^*$ enters in Eq.~\eqref{eq:eir_by_2}. The two masses ($m^*,M^*$) correspond to the
cyclotron motion and the guiding center motion, respectively. We avoid
the use of a Dingle temperature \cite{Jain}.

\item[(B)] The background resistance $\langle R \rangle$ averaged over the
field $B$ is zero:
\begin{equation}
\langle R \rangle \equiv \left\langle \frac{V}{I} \right\rangle = 0.
\label{eq:tfi_zero}
\end{equation}
This behaviour is in agreement with formula \eqref{eq:cWt_obtain_2}. It arises from the
fact that there is no Landau-like term proportional to the squared magnetic field $B^2$
in the statistical weight $W$ in 2D, see Eq.~\eqref{eq:Wt_obtain}. (There is no Landau
diamagnetism in 2D in contrast to the 3D case.)

\item[(C)] The flat minimum meaning zero resistance $R = 0$, indicates the
existence of a superconducting state. The superconducting state is
stable with an energy gap. The supercurrents run with no scatterings by
impurities and phonons. As is well known, the magnetic field is
detrimental to the superconducting state. If the excess magnetic field
$B^*$ relative to the center field of the horizontal stretch exceeds a
critical field, then the superconductivity is destroyed. The microscopic
origin of this effect was explained in section 2. Briefly, the
supercurrent is composed of the positively and negatively charged
pairon-currents. The excess magnetic field $B^*$ generates oppositely
directed forces and breaks up $\pm$ pairons (Cooper pairs).

\item[(D)] The magnetic field energy is quadratic in the excess field $B^*$,
see Eq.~\eqref{eq:fes_that}, which explains the right-left symmetry of the destroyment
of the superconducting state.

\item[(E)] The integer QHE occurring at the LL occupation numbers $\nu = P =
1,2,\dots$, have the quantized magnetic fluxes:
\begin{equation}
BA = \frac{1}{P} \Phi_0 N_{\phi_1} = \frac{1}{P}\left( \frac{h}{e} \right) N_{\phi_1},
\label{eq:quantized_magnetic_fluxes}
\end{equation}
where $N_{\phi_1}$ is the fluxon number at $\nu = P = 1$. Hence the QHE
has maxima at
\begin{equation}
B = \frac{1}{P}\left( \frac{h}{e} \right) n_{\phi_1}, \quad n_{\phi_1} \equiv \frac{N_{\phi_1}}{A}.
\label{eq:Qhm_at}
\end{equation}
Equations \eqref{eq:cWt_obtain_3} and \eqref{eq:bwo_as} indicate that the resistivity $r$ has minima when
\begin{equation}
\cos (2 \pi \varepsilon_\text{F} / \hbar \omega_c) = 1,
\label{eq:rhm_when}
\end{equation}
whose solutions are
\begin{equation}
\frac{2 \pi \varepsilon_\text{F}}{\hbar \omega_c} = 2 \pi Q,\quad Q = 1,2,\dots.
\label{eq:ws_are}
\end{equation}
We use $\omega_c = eB / m^*$, $\varepsilon_\text{F} =
p_\text{F}^2 / 2m^*$, $2 \pi p_\text{F}^2 = n_e e$, and solve Eq.~\eqref{eq:ws_are}
for $B$ and obtain
\begin{equation}
B = \frac{h}{e} \frac{1}{Q} n_{e_1},
\label{eq:Efa_obtain}
\end{equation}
where $n_{e_1}$ is the electron density at $\nu = 1$.
From Eq.~\eqref{eq:tbd_over}, we obtain
\begin{equation}
n_{e_1} = n_{\phi_1}.
\label{eq:FEw_obtain}
\end{equation}
Both $P$ and $Q$ are positive integers. Hence we find from
Eqs.\ \eqref{eq:Qhm_at} and \eqref{eq:Efa_obtain} that the integer QHE
maxima and the SdH minima occur precisely at the same magnetic fields
$B$. Thus, the QHE minima progressively turn into the SdH maxima with
increasing DC.
\end{enumerate}

\begin{figure}
\centering
\includegraphics{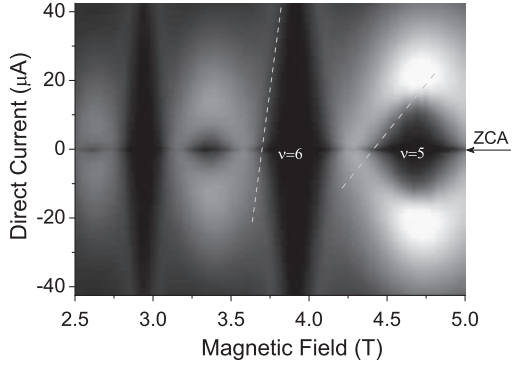}
\caption{The differential resistance $r$ of an InGaAs/InP Hall bar ($w =
  100\,\mu$m) is plotted versus magnetic field and DC. The ZCA position
  is indicated by arrow.}\label{fg:Figure4}
\end{figure}

Fig.~2, Ref.~\cite{Studenikin} is reproduced in Fig.~\ref{fg:Figure4}.%
The differential resistance $r \equiv dV/dI$ is plotted versus magnetic field (T) and direct current
($\mu$\,A). Diamond-shaped regions near SdH minima are called
transport diamonds (TD).

Our interpretation of the TD is a break-down of the superconducting QH
state due to the excess magnetic field and the direct current. The
direct current by itself generates a magnetic field, which is
detrimental to the superconducting state.

\begin{figure}
\centering
\includegraphics{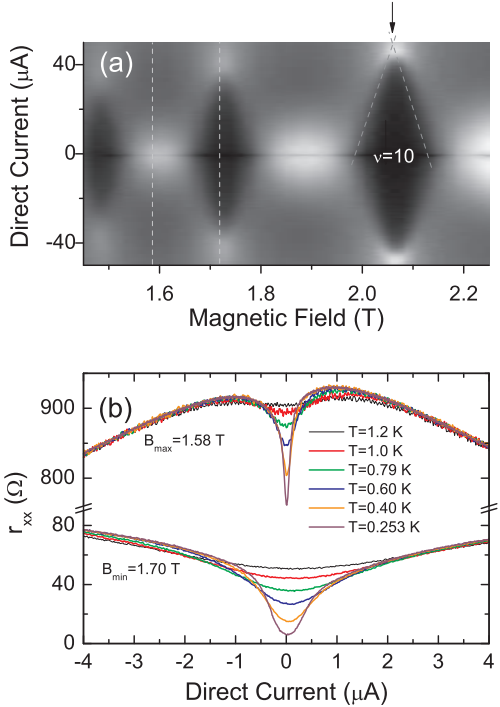}
\caption{(a) The differential resistance of an InGaAs/InP Hall bar ($w =
  100\,\mu$m), $T=300$\,mK, is plotted versus magnetic field and
  DC; (b) the ZCA at two magnetic fields ($1.58$\,T, $1.70$\,T) indicated
  by vertical dashed lines in (a) for different temperatures.}\label{fg:Figure5}
\end{figure}

Fig.~\ref{fg:Figure5} is reproduced after Ref.~\cite{Studenikin}, Fig.~3.%
In (a) transport diamonds are shown, which are similar to those in
Fig.~\ref{fg:Figure2}. The temperature-dependence of the differential
resistance is shown in (b) in the range ($0.253$--$1.2$\,K).

The sharp dip in $r$ vs.\ DC near $\text{DC} = 0$ observed in Fig.~\ref{fg:Figure3} and
Fig.\ \ref{fg:Figure5}(a) is called the ZCA, the narrow horizontal line indicated by an
arrow in Fig.~\ref{fg:Figure3}. Its temperature behavior at $B = 1.58$\,T and $B
= 1.70$\,T is shown in Fig.\ \ref{fg:Figure5}(b).

The original authors \cite{Studenikin} suspect that the origin of the ZCA arises from the
Coulomb gap in the one particle density of states of interacting
electrons. We propose a differing interpretation:

Let us consider the case of the SdH at $B = 1.58$\,T, the top
figure in Fig.\ \ref{fg:Figure5}(b). At the low temperatures $T =
(0.253$--$1.2)$\,K, the optical phonon population given by the
Plank distribution function can be approximated by the Boltzmann
distribution function:
\begin{equation}
n_\text{ph} = \frac{1}{e^{\beta \varepsilon_0}-1} \cong e^{- \varepsilon_0/k_\text{B}T},
\label{eq:Boltzmann_distribution_function}
\end{equation}
where $\varepsilon_0$ is the longitudinal optical phonon energy (assumed
constant) and $k_\text{B}$ the Boltzmann constant. The temperature
dependence is exponential. The phonon population $n_\text{ph}$ is
rapidly changing with temperature and dominates. The resistance $R$ is
proportional to the electron-phonon scattering rate $\gamma_\text{ph}$:
\begin{equation}
R \equiv \sigma^{-1} \propto \gamma_\text{ph} = n_\text{ph} v_\text{ph} \mathcal{A},
\label{eq:tes_rate}
\end{equation}
where $v_\text{ph}$ is the phonon speed, $\mathcal{A}$ the
electron-phonon scattering cross section, and $n_\text{ph}$ the phonon
population given in Eq.~\eqref{eq:Boltzmann_distribution_function}.

Studenikin \emph{et al}.\ \cite{Studenikin} observed that the temperature dependence of
the ZCA follows the Arrhenius law:
\begin{equation}
\gamma \propto e^{- \varepsilon_A / k_\text{B}T}
\label{eq:Arrhenius_law}
\end{equation}
with the activation energy
\begin{equation}
\varepsilon_A / k_\text{B} = 1.3\,\text{K}.
\label{eq:activation_energy}
\end{equation}
This value may correspond to the optical phonon energy $\varepsilon_0$:
\begin{equation}
\varepsilon_A = \varepsilon_0.
\label{eq:optical_phonon_energy}
\end{equation}
This finding supports our view that the temperature dependence of the ZCA
arises from the electron-phonon scattering.

We next consider the ZCA for the QHE at $B = 1.70$\,T. This ZCA
is also temperature-dependent. As the temperature decreases from $1.2$\,K to $0.253$\,K, the negative peak decreases in
magnitude and its width becomes narrower. In the QHE under radiation, a
supercurrent due to moving pairons condensed run in the upper (excited)
channel and a normal current due to electrons run in the base channel.
The resistance of the normal current is proportional to the
electron-phonon scattering rate $\gamma_\text{ph}$, as shown in Eq.~\eqref{eq:Qhm_at}. Then,
the phonon population approximately decreases exponentially at low
temperatures (below $1.2$\,K). Thus the resistance decreases as
the temperature $T$ is lowered, which explains the observed temperature
dependence.

The QHE at zero DC is destroyed either by increasing excess magnetic
fields or by increasing DC-induced magnetic fields. But the ZCA
indicates the destroyment is sharper for the case of increasing DC. This
difference should arise from the direction of the magnetic field. The DC
running along the sample length is likely to be inhomogeneous, stronger
at the outer edge. Then the superconductivity is destroyed at the edges first
according to Silsbeeb rule. On the other hand, the applied magnetic
field alone should keep the current homogeneous.

Studenikin \emph{et al}.\ \cite{Studenikin} observed essentially same TD
and ZCA in heterojunction GaAs/AlGaAs. In particular, the QHE minima
progressively turn to the SdH maxima as DC increases, and the ZCA is
sharp near $\text{DC} = 0$. The same theory applies here. The authors
thank Dr.\ S.~Studenikin for enlightening discussions.

\appendix

\makeatletter
\def\@seccntformat#1{Appendix\ \csname the#1\endcsname\quad}
\makeatother

\section{DERIVATION OF EQS.~(\ref{eq:eao_from}) AND (\ref{eq:soE_as})}\label{ap:Derivation_A}

Dropping the ``holes'' from the Hamiltonian $\mathcal{H}$ in Eq.~\eqref{eq:tpv_eliminated},
we obtain
\begin{align}
\mathcal{H}_c = &\sum_\mathbf{k} \sum_\mathbf{q}
\left( \varepsilon_{|\mathbf{k}\boldsymbol{+}\mathbf{q}/2|} +
\varepsilon_{|\boldsymbol{-}\mathbf{k}\boldsymbol{+}\mathbf{q}/2|}^{(3)} \right)
B_\mathbf{kq}^\dagger B_\mathbf{kq} \nonumber\\
&- v_0 \sideset{}{'}{\sum}_\mathbf{q} \sideset{}{'}{\sum}_\mathbf{k} \sideset{}{'}{\sum}_{\mathbf{k}\boldsymbol{'}} B_{\mathbf{k}\boldsymbol{'}\mathbf{q}}^\dagger B_\mathbf{kq},
\label{eq:iEw_obtain}
\end{align}
where we suppressed the ``electron'' and spin indices. Using the
anticommutation rules \eqref{eq:soE_as}, we obtain
\begin{align}
\left[ \mathcal{H}_c, B_\mathbf{kq}^\dagger \right] = &\left( \varepsilon_{|\mathbf{k}\boldsymbol{+}\mathbf{q}/2|} + \varepsilon_{|\boldsymbol{-}\mathbf{k}\boldsymbol{+}\mathbf{q}/2|}^{(3)} \right)
B_\mathbf{kq}^\dagger \nonumber\\
&\mbox{}- v_0\sideset{}{'}{\sum}_{\mathbf{k}\boldsymbol{'}}
B_{\mathbf{k}\boldsymbol{'}\mathbf{q}}^\dagger\left( 1 -
n_{\mathbf{k}\boldsymbol{+}\mathbf{q}/2} -
n_{\boldsymbol{-}\mathbf{k}\boldsymbol{+}\mathbf{q}/2}^{(3)} \right).
\label{eq:arw_obtain}
\end{align}
The Hamiltonian $\mathcal{H}_c$ is bilinear in $(B,B^\dagger)$, and can
therefore be diagonalized exactly:
\begin{equation}
\mathcal{H}_c = \sum_\mu w_\mu \phi_\mu^\dagger \phi_\mu,
\label{eq:tbd_exactly}
\end{equation}
where $w_\mu$ is the energy and $\phi_\mu$ the annihilation operator. We
multiply Eq.~\eqref{eq:arw_obtain} by $\phi_\mu$ from the right, take a grand canonical
ensemble average, denoted by angular brackets, and get
\begin{multline}
w_\mu \Psi_\mu(\mathbf{k,q}) = \left(
\varepsilon_{|\mathbf{k}\boldsymbol{+}\mathbf{q}/2|} +
\varepsilon_{|\boldsymbol{-}\mathbf{k}\boldsymbol{+}\mathbf{q}/2|}^{(3)}
\right) \Psi_\mu(\mathbf{k,q})\\
- \frac{v_0}{(2\pi\hbar)^2}\int' d^2k'
\Psi_\mu(\mathbf{k',q})\\
\times \Big\langle 1 - f_\text{F} \left(
\varepsilon_{|\mathbf{k}\boldsymbol{'+}\mathbf{q}/2|} \right)
- f_\text{F} \left( \varepsilon_{|\boldsymbol{-}\mathbf{k}\boldsymbol{'+}\mathbf{q}/2|}^{(3)} \right)
\Big\rangle,
\label{eq:aba_get}
\end{multline}
where $\langle n_p \rangle = f_\text{F} (\varepsilon_p)$ is the Fermi
distribution function. The reduced wavefunction
\begin{equation}
\Psi_\mu(\mathbf{k,q})
\equiv \left\langle B_\mathbf{kq}^\dagger \phi_\mu \right\rangle =
\langle \boldsymbol{\mu} |\hat{n}| \mathbf{k,q} \rangle
\label{eq:reduced_wavefunction}
\end{equation}
can be regarded as the mixed representation of the \emph{reduced}
density operator $\hat{n}$ defined through $\langle
\mathbf{k}\boldsymbol{'},\mathbf{q}\boldsymbol{'} |\hat{n}| \mathbf{k,q}
\rangle \equiv \left\langle B_\mathbf{k,q}^\dagger
B_{\mathbf{k}\boldsymbol{'},\mathbf{q}\boldsymbol{'}} \right\rangle$.
The fc-boson energy $w_\mu$ can be specified by $(N_\text{L},q)$, and it
will be denoted by $w_q$ since it is $N_\text{L}$-independent. As $T
\rightarrow 0$, $f_\text{F} (\varepsilon_p) \rightarrow 0$. Dropping the
fluxon energy and replacing $\mathbf{q}/2$ by $\mathbf{q}$, we obtain
Eq.~\eqref{eq:eao_from}. We solve this equation, assuming $\varepsilon_\text{F} \gg
\hbar \omega_\text{D}$. Using a Taylor series expansion, we obtain
Eq.~\eqref{eq:soE_as} to the linear in $q$.

\section{DERIVATION OF EQ.~(\ref{eq:critical_temperature})}\label{ap:Derivation_B}

The BEC occurs when the chemical potential $\mu$ vanishes at a finite
$T$. The critical temperature $T_c$ can be determined from
\begin{equation}
n = (2\pi\hbar)^{-2} \int d^2 p \left[e^{\beta_c \varepsilon} - 1\right]^{-1},\quad
\beta_c \equiv (k_\text{B} T_c)^{-1}.
\label{eq:cbd_from}
\end{equation}
After expanding the integrand in powers of $e^{-\beta_c \varepsilon}$
and using $\varepsilon = cp$, we obtain
\begin{equation}
n = 1.654 (2\pi)^{-1} (k_\text{B} T_c / \hbar c)^2,
\label{eq:auw_obtain}
\end{equation}
yielding Eq.~\eqref{eq:critical_temperature}.

\section{STATISTICAL WEIGHT FOR THE LANDAU STATES}\label{ap:SW_LS}

The statistical weight $W$ for the Landau states in 2D will be
calculated in this appendix. We write the sum in Eq.~\eqref{eq:Hw_obtain} as
\begin{equation}
2 \sum_{n=0}^\infty \Theta (\epsilon - (2n + 1)\pi) = \Theta (\epsilon -
\pi) +\psi (\epsilon; 0),
\label{eq:siE_as}
\end{equation}
\begin{equation}
\psi(\epsilon; x) \equiv \sum_{n = -\infty}^\infty \Theta (\epsilon - \pi
- 2\pi |n + x|).
\label{eq:siE_as_2}
\end{equation}
Note that $\psi(\epsilon; x)$ is periodic in $x$ and can therefore be
expanded in a Fourier series. After the Fourier expansion, we set $x=0$
and obtain Eq.~\eqref{eq:siE_as}. By taking the real part (Re) of Eq.~\eqref{eq:siE_as} and
using Eq.~\eqref{eq:rta_obtain}, we obtain
\begin{multline}
\text{Re\{Equation\ \eqref{eq:siE_as}\}} = \frac{1}{\pi} \int_0^\infty d\tau \Theta (\epsilon
- \tau)\\
+ \frac{2}{\pi} \sum_{\nu=1}^\infty (-1)^\nu \int_0^\infty d\tau
\Theta (\epsilon - \tau) \cos \nu \tau,
\label{eq:uEw_obtain}
\end{multline}
where we assumed $\epsilon \equiv 2\pi E / \hbar \omega_c \gg 1$ and
neglected $\pi$ against $\epsilon$. The integral in the first term in
Eq.~\eqref{eq:uEw_obtain} yields $\epsilon$. The integral in the second term is
\begin{equation}
\int_0^\infty d\tau \Theta (\epsilon - \tau) \cos \nu \tau =
\frac{1}{\nu} \sin \nu \epsilon.
\label{eq:tst_is}
\end{equation}
We then obtain
\begin{equation}
\text{Re\{Equation\ \eqref{eq:siE_as}\}} = \frac{1}{\pi}\epsilon + \frac{2}{\pi}\sum_{\nu =
  1}^\infty \frac{(-1)^\nu}{\nu} \sin \nu \epsilon.
\label{eq:Wt_obtain_4}
\end{equation}
Using Eqs.\ \eqref{eq:rta_obtain} and \eqref{eq:Wt_obtain_4}, we obtain
\begin{align}
W(E) &= W_0 + W_\text{osc} \nonumber\\
     &= C\hbar \omega_c \left( \frac{\epsilon}{\pi} \right) + C \hbar \omega_c
\frac{2}{\pi} \sum_{\nu = 1}^\infty \frac{(-1)^\nu}{\nu} \sin \left(
\frac{2\pi\nu E}{\hbar \omega_c} \right).
\label{eq:Eaw_obtain}
\end{align}

\section{DERIVATION OF EQ.~(\ref{eq:integral_formula})}\label{ap:Derivation_C}

Let us consider an integral on the real axis
\begin{equation}
I(y,\alpha,R) = \int_{-R}^Rdx \frac{e^{\text{i}\alpha (x +
 \text{i}y)}}{e^z + 1},\quad z = x + \text{i}y \text{ and } \alpha,R > 0.
\label{eq:otr_axis}
\end{equation}
We add an integral over a semicircle of radius $R$ in the upper
$z$-plane to form an integral over a closed contour. We then take the
limit: $R \rightarrow \infty$. The integral over the semicircle vanishes
in this limit if $\alpha > 0$. The integral on the real axis,
$I(y,\alpha,\infty)$, becomes the desired integral in Eq.~\eqref{eq:integral_formula}. The
integral over the closed contour can be evaluated by using the residue
theorem. Note that $(e^z + 1)^{-1}$ has simple poles at $z = \pi \text{i},3\pi \text{i},\dots,(2n-1)\pi \text{i},\dots.$
We may use the following formula valid for a simple pole at $z = z_j$:
\begin{equation}
\text{Res}\{p(z)/q(z), z_j\} = p(z_j)/q'(z_j),
\label{eq:asp_at}
\end{equation}
where $p(z)$ is analytic at $z = z_j$, and the symbol Res means a
residue. We then obtain
\begin{align}
I(\alpha,\infty) &= 2\pi \text{i} \cdot \sum_{n=1}^\infty \text{Res} \left\{
\frac{e^{\text{i}\alpha z}}{e^z + 1}, z_n = (2n - 1) \pi \text{i} \right\} \nonumber\\
&= 2\pi \text{i} \cdot \sum_{n=1}^\infty \frac{e^{\text{i}\alpha [(2n - 1) \pi \text{i}]}}{e^{(2n - 1) \pi \text{i}}}
= - 2\pi \text{i} \frac{e^{- \alpha \pi}}{1 - e^{- 2 \alpha \pi}} \nonumber\\
&= \frac{\pi}{\text{i}}\frac{1}{\sinh \alpha \pi}.
\label{eq:rWt_obtain}
\end{align}

\end{document}